\documentclass[aps,pre,nofootinbib,amsfonts,superscriptaddress,longbibliography,showkeys,notitlepage]{revtex4-1}
\usepackage{hyperref}
\usepackage{graphicx,bm,amsmath,color}
\usepackage[latin1]{inputenc}
\usepackage[english]{babel}
\usepackage{bm}
\usepackage[T1]{fontenc}
\usepackage{epstopdf}
\usepackage{booktabs}
\usepackage{placeins}
\usepackage{multirow}
\usepackage[caption=false]{subfig}
\newcommand{\ba}{\begin{array}}
\newcommand{\ea}{\end{array}}

\begin{document}

\title{Directionality reduces the impact of epidemics in multilayer networks}

\author{Xiangrong Wang}
\affiliation{Faculty of Electrical Engineering, Mathematics and Computer Science, Delft University of Technology, The Netherlands}
\thanks{X.W. and A.A. contributed equally to this work.}
\author{Alberto Aleta}
\email{albertoaleta@gmail.com}
\affiliation{Institute for Biocomputation and Physics of Complex Systems, University of Zaragoza, Zaragoza, Spain}
\affiliation{Department of Theoretical Physics, University of Zaragoza, Spain}
\thanks{X.W. and A.A. contributed equally to this work.}
\author{Dan Lu}
\affiliation{Institute for Biocomputation and Physics of Complex Systems, University of Zaragoza, Zaragoza, Spain}
\affiliation{Department of Theoretical Physics, University of Zaragoza, Spain}
\author{Yamir Moreno}
\email{yamir.moreno@gmail.com}
\affiliation{Institute for Biocomputation and Physics of Complex Systems, University of Zaragoza, Zaragoza, Spain}
\affiliation{Department of Theoretical Physics, University of Zaragoza, Spain}
\affiliation{ISI Foundation, Turin, Italy}


\begin{abstract}
The study of how diseases spread has greatly benefited from advances in network modeling. Recently, a class of networks known as multilayer graphs have been shown to describe more accurately many real systems, making it possible to address more complex scenarios in epidemiology such as the interaction between different pathogens or multiple strains of the same disease. In this work, we study in depth a class of networks that have gone unnoticed up to now, despite of its relevance for spreading dynamics. Specifically, we focus on directed multilayer networks, characterized by the existence of directed links, either within the layers or across layers. Using the generating function approach and numerical simulations of a stochastic susceptible-infected-susceptible (SIS) model, we calculate the epidemic threshold for these networks for different degree distributions of the networks. Our results show that the main feature that determines the value of the epidemic threshold is the directionality of the links connecting different layers, regardless of the degree distribution chosen. Our findings are of utmost interest given the ubiquitous presence of directed multilayer networks and the  widespread use of disease-like spreading processes in a broad range of phenomena such as diffusion processes in social and transportation systems.
\end{abstract}

\maketitle

Directionality in contact networks has often been disregarded, either because of the lack of data or in order to simplify theoretical approaches \cite{Newman2018Jul}. This is the case of disease spreading models, which usually consider the underlying networks as undirected \cite{Pastor-Satorras2015Aug, deArruda2018Oct}. However, there are scenarios in which directionality has been found to be a key feature. Relevant examples are the case of meerkats in which transmission varies between groomers and groomees \cite{Drewe2010Feb} and the transmission of HIV between humans, with male-to-female transmission being 2.3 times greater than female-to-male transmission \cite{Nicolosi1994Nov}. Similarly, when addressing the problem of diseases that can be transmitted among different species, it is important to account for the fact that they might be able to spread from one type of host to the other, but not the other way around. For example, the bubonic plague can be endemic in rodent populations and spread to humans and other animals under certain conditions. If it evolves to the pneumonic form, it may then spread from human to human \cite{Kool2005Apr}. Analogously, Andes virus usually spreads within rodent populations, but it can be transmitted to humans and then spread via person-to-person contacts \cite{Martinez2005Dec}. These types of interspecies contagions and other similar cases can be studied using multilayer networks, in which the network of each species is encoded in the layers and the possible interspecies interactions are given by the links that connect the layers \cite{Aleta2018Oct}. 

The use of directed multilayer networks is not constrained to diseases that can infect human populations. Indeed, analogous scenarios can be found in the interface between wildlife and livestock, with diseases being endemic in one of them and then being transmitted unidirectionaly to the other \cite{Bengis2002Apr}. This directionality is particularly relevant in the surveillance of diseases within the livestock industry, where the direction of the livestock interchange between farms can uncover structural changes that would be otherwise hidden \cite{Bajardi2011May}. Even more, the recent introduction of high resolution data of face-to-face interactions has also renewed the interest in using directed networks both in human and animal populations \cite{Genois2018Dec,dogs2018}. This data can be used to build temporal multilayer networks in which the connections between layers, i.e., different time frames, have to be necessarily directed in order to preserve the causality induced by time ordering \cite{Valdano2015Apr}. 

In this work, we aim at characterizing the spreading of diseases in directed multiplex networks. We focus on investigating how the epidemic threshold is influenced by the directionality of both interlayer and intralayer links (see \emph{Materials and Methods}). In particular, we consider multiplex networks composed by two layers with either homogeneous or heterogeneous degree distributions in the layers. Besides, we analyze several combinations of directionality: (i) Directed layer - Undirected interlinks - Directed layer (DUD); (ii) Directed layer - Directed interlinks - Directed layer (DDD); and (iii) Undirected layer - Directed interlinks - Undirected layer (UDU). For the sake of comparison, we also include the standard scenario, namely, (iv) Undirected layer - Undirected interlinks - Undirected layer (UUU). We then implement a susceptible-infected-susceptible (SIS) model on these networks and study the evolution of the epidemic threshold as a function of the directionality and the coupling strength between layers. In addition, we analytically derive the epidemic thresholds using generating functions, which allows to provide theoretical insights on the underlying mechanisms driving the dynamics of these systems. Our results show that the presence of directed links results in larger epidemic thresholds with respect to the case of undirected networks, and that the system is more resilient when the interlayer links are directed.

\section{Results\label{section:results}}

We first present results of numerical simulations of a stochastic susceptible-infected-susceptible model. In this model, nodes can be either susceptible or infected. The latters spread the disease to the formers if they are in contact with a given probability. One of the main characteristics of multiplex networks is the existence of several types of links. Thus, it is possible to associate different spreading probabilities to each of these links \cite{PhysRevX.7.011014}. In our model, we assume two spreading probabilities: the interlayer spreading probability, $\gamma$, and the intralayer spreading probability, $\beta$. Hence, an infected node transmits the disease with probability $\beta$ to those susceptible neighbors of the same layer and with probability $\gamma$ to those located in other layers (for further details see \emph{Materials and Methods}, section \ref{mat:stochastic}). This distinction implies that it is possible to find a critical value of $\beta$ for each value of $\gamma$ and vice versa. Thus, henceforth we will define the epidemic threshold as  $\beta_c$ and explore its value as a function of $\gamma$.

\begin{figure*}
\centering
\includegraphics[width=0.8\textwidth]{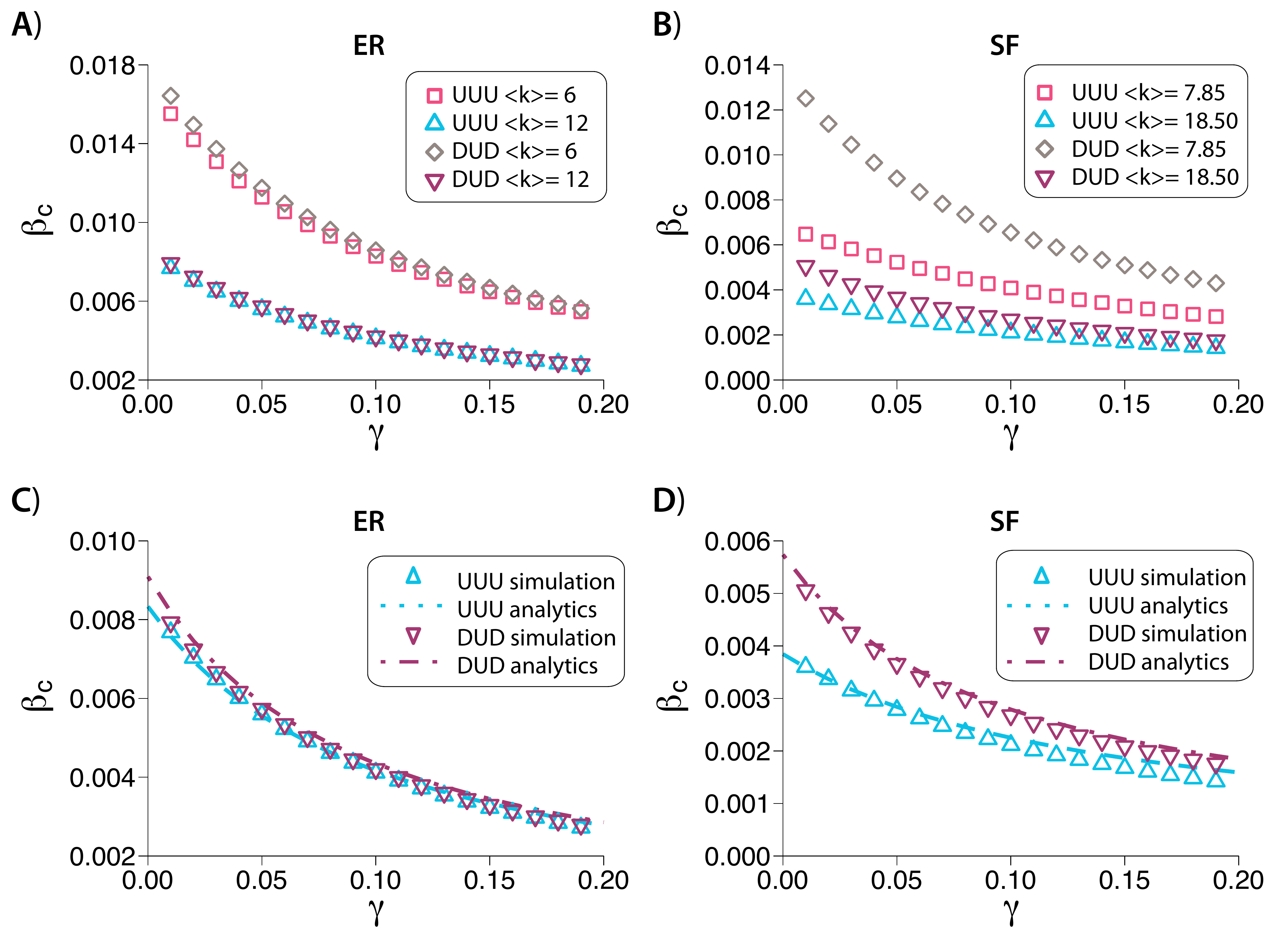}%
\caption{Epidemic threshold for the spread of a disease within layers, $\beta_c$, as a function of the probability of interlayer contagion, $\gamma$. Panels (A) and (B) show results for the UUU and DUD configurations with ER (A) and SF (B) degree distributions in the layers. In all cases $\mu=0.1$, the number of nodes is $N=2\cdot 10^4$ and for each directionality configuration there are two sets of networks: in the ER case one with $\langle k \rangle = 6$ in both layers and another one with $\langle k \rangle = 12$ in both layers; in the SF case one with $k_\text{min}=4 \text{ and } \alpha=2.7$ (average degree $\langle k \rangle = 7.85$) and another one with ${k_\text{min}}=10 \text{ and } \alpha=2.8$ (average degree $\langle k \rangle = 18.50$). In panels (C) and (D) we compare the analytical values of $\beta_c$ with corresponding results from the numerical simulations for the same networks and directionality configurations shown in panels (A) and (B).}%
\label{fig:UUU_DUD}%
\end{figure*}

The SIS dynamics is implemented on directed multiplex networks composed by two layers. As previously stated, we explore four different configurations of directionality denoted as DUD, DDD, UDU and UUU. Furthermore, to define the degree distribution in the layers we use power-law and Poisson distributions, which correspond respectively to Scale-Free (SF) and Erd\H{o}s-Rényi (ER) network models. In figure \ref{fig:UUU_DUD} we show the evolution of the epidemic threshold, $\beta_c$, as a function of $\gamma$ for the configurations with undirected interlinks, UUU and DUD, both for ER (\ref{fig:UUU_DUD}A) and SF (\ref{fig:UUU_DUD}B) networks, for two different average degrees $\langle k \rangle$.

\begin{figure*}
\centering
\includegraphics[width=17.8cm]{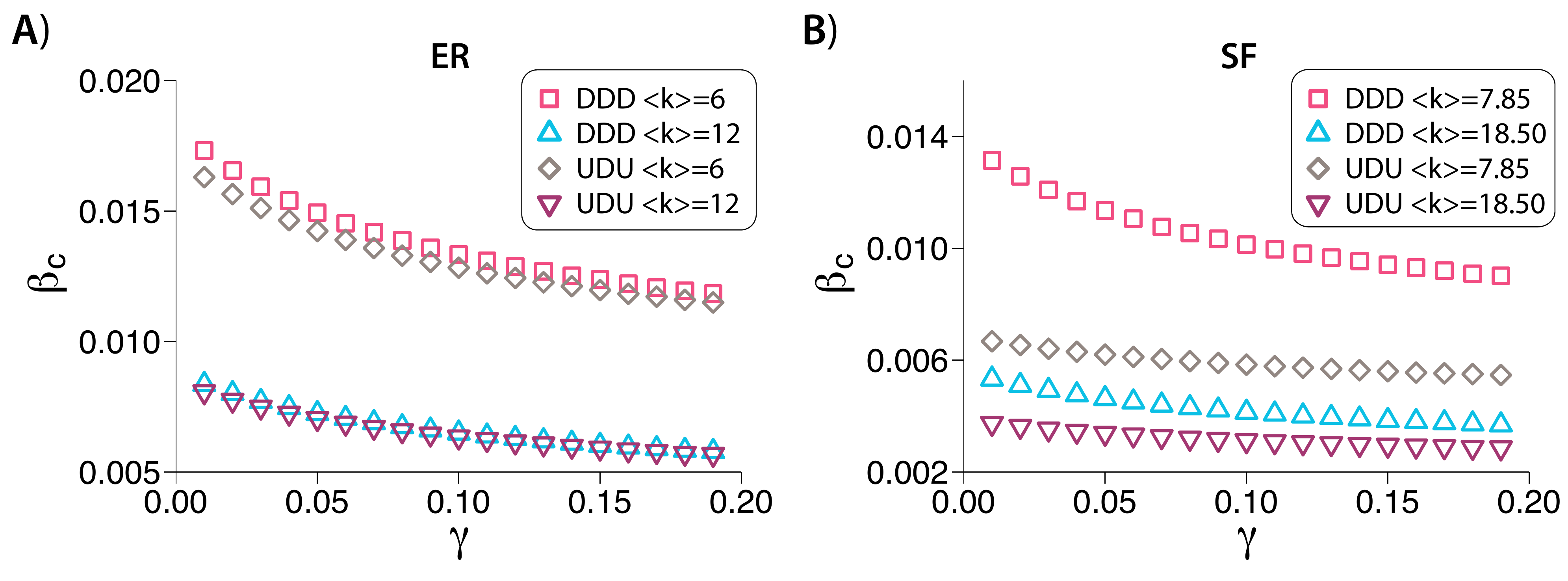}%
\caption{Critical value of the within-layer spreading rate, $\beta_c$, as a function of the spreading rate across layers, $\gamma$, in DDD and UDU configurations built up using the $p$-model with ER (A) and SF (B) degree distributions in the layers. In all cases $p=0.5$, $\mu=0.1$, the number of nodes is $N=2\cdot 10^4$ and for each directionality configuration there are two sets of networks: in the ER case one with $\langle k \rangle = 6$ in both layers and another one with $\langle k \rangle = 12$ in both layers; in the SF case one with $k_\text{min}=4 \text{ and } \alpha=2.7$ (average degree $\langle k \rangle = 7.85$) and another one with  ${k_\text{min}}=10 \text{ and } \alpha=2.8$ (average degree $\langle k \rangle = 18.50$). The dependence of the results with $p$ is presented in \emph{Supplementary Information}, figure \emph{S3}.}%
\label{fig:ER_DDD_UDU}%
\end{figure*}

For the cases in which the interlinks are directed, we need to define how many links point from one layer $u$ to another layer $v$, either in the $u\rightarrow v$ direction or in the opposite one, $u \leftarrow v$. Indeed, if we set all interlinks to have the same direction, the epidemic threshold would be trivially the one of the source layer and thus the multiplex structure would play no role. For this reason, for each directed link connecting layers $u$ and $v$ we set the directionality to be $u \rightarrow v$ with probability $p$ and $u \leftarrow v$ with probability $(1-p)$. Consequently, in networks with directed interlinks the epidemic threshold will be given as a function of this probability $p$. We refer to this procedure of generating interlinks as the $p$-model. The same dependence of the critical threshold depicted in Figure \ref{fig:UUU_DUD} is shown in Figure \ref{fig:ER_DDD_UDU} for DDD and UDU configurations built using the $p$-model. 

It is also possible to study scenarios in which each interlink does not only have one possible directionality, either $u\rightarrow v$ or $u \leftarrow v$, but instead are bi-directional. This is achieved by setting two independent probabilities $-$one for each direction $-$, thus allowing for the coexistence of single directionality and bi-directionality in the interlinks. This situation, which we denote as the $pq$-model, is further analyzed in the \emph{Supplementary Information}, section 1. 

Lastly, in order to obtain insights into the mechanisms behind driving the spreading process on the directed multiplex networks, we analytically derive the epidemic threshold for all the configurations considered in this work, both for ER and SF networks. To this end, we extend the generating function formalism, which has been used previously in the context of directed monolayer networks \cite{Meyers2006Jun} and interdependent directed networks \cite{Liu1138}, to multiplex networks. This formalism is outlined in the next subsection \ref{sec:generating}. The complete derivation of the epidemic threshold is presented in the \emph{Supplementary Information}, section 1.

\subsection{Generating function\label{sec:generating}} Within the generating function formalism, a node has an in-degree  $j$, out-degree $k$ and inter-degree $m$ with probability $p_{jlm}$, being the first two related to the links contained in each layer and the latter to links connecting nodes in different layers. The generating function for the degree distribution of a node is then defined as
\begin{equation}\label{eq:degree distribution}
G(x,y,z) = \sum_{j=0}^\infty \sum_{l=0}^\infty \sum_{m=0}^\infty p_{jlm} x^j y^{l} z^{m}
\end{equation}

\noindent
so that in order to describe a particular network it is sufficient to set $p_{jlm}$ to the degree distribution of the network. Indeed, with this function it is possible to characterize several properties of the network such as the excess degree which is the main quantity that is needed for the derivation of the epidemic threshold. The excess degree of a node is defined as the number of links of a node reached by following a randomly chosen link, without including the incoming link. Hence, the distribution of excess degree of a node that is reached by following a directed link in its direction is generated by
\begin{equation}\label{eq:Hd}
H_d (x,y,z) = \frac{1}{\langle k_d \rangle } G^{(1,0,0)}(x,y,z)
\end{equation}

\noindent
where $\langle k_d\rangle $ is the average directed degree and the superscript $(1,0,0)$ refers to partial derivation with respect to $x$. Similar expressions can be obtained for the excess degree of a node reached via the reverse direction of the same directed link and via an undirected link (see \emph{Supplementary Information}, section 1).

The size of an outbreak, as well as the epidemic threshold, can be obtained by computing the fraction of occupied links in the network. In this context, occupied link refers to a link through which the disease was transmitted. This can be accounted for by incorporating the transmissibility, i.e., the mean probability of transmission between individuals \cite{Newman2002Jul}, to the previous equations so that
\begin{equation}\label{eq:degree_distribution_T}
G(x,y,z;T,T_{uv}) =  \\  G(1-T+Tx,1-T+Ty,1-T_{uv}+T_{uv}z)
\end{equation}
\noindent
where $T$ and $T_{uv}$ denote the transmissibility within a layer and across layers, respectively. Recalling that $\beta$ is the within-layer transmission rate, that $\gamma$ is the transmission rate through links that connect different layers and that $\mu$ is the recovery rate, these quantities can be expressed as (see \emph{Supplementary Information}, section 1F):
\begin{equation}\label{eq:T}
T = 1 - \frac{\mu}{\beta + \mu}
\end{equation}
\begin{equation}\label{eq:Tuv}
T_{uv} = 1 - \frac{\mu}{\gamma + \mu}
\end{equation}

The generating function for the distribution of the size of an outbreak can be expressed as
\begin{equation}\label{eq:g}
g(w;T,T_{uv}) = wG(1,h_1(w;T,T_{uv}),h_{12}(w;T,T_{uv});T,T_{uv})
\end{equation}

\noindent
where $h_1$ and $h_{12}$ are recursive functions that generate the distribution of the size of an outbreak starting at a link connecting nodes in layer 1  and at a link connecting nodes in layer 1 and 2 respectively. The average size of an outbreak will be then given by the derivative with respect to $w$ of $g(w;T,T_{uv})$ evaluated at $w=1$. The said derivative goes to infinity when its denominator equals 0, which characterizes a phase transition from a phase in with only small size outbreaks to one characterized by the occurrence of macroscopic outbreaks. Thus, the epidemic threshold can be obtained from the equality

\begin{equation}\label{eq:denominator}
\begin{split}
&\left[\left(1-H_{1}^{(0,1,0)}\right)H-H_{1}^{(0,0,1)}H_{12}^{(0,0,1)}H_{21}^{(0,1,0)}\right]\\
&\left[\left(1-H_{2}^{(0,1,0)}\right)H-H_{2}^{(0,0,1)}H_{21}^{(0,0,1)}H_{12}^{(0,1,0)}\right]\\
&-H_{1}^{(0,0,1)}H_{2}^{(0,0,1)}H_{12}^{(0,1,0)}H_{21}^{(0,1,0)}=0
\end{split}
\end{equation}

\noindent
where $ H = 1-H_{12}^{(0,0,1)}H_{21}^{(0,0,1)}$ and $H_i$ refers to transmission within layer $i$ and $H_{ij}$ to transmission from layer $i$ to layer $j$. 

The above expression is general enough as to be used in the calculation of the epidemic threshold for each of the cases considered in this work. To this end, the only step that is left is to substitute $H_i$ by $H_d$ if the links in layer $i$ are directed or, conversely, by $H_u$ if they are undirected (see \emph{Supplementary Information}, section 1 and figures \emph{S1-S2}). In what follows, we present the results obtained for the thresholds after considering directionality (or lack thereof) and different network topologies.

\subsection{ER networks\label{section:ER}}

In ER networks the degree distribution follows a Poisson distribution. If we consider an UUU network with nodes in both layers following said degree distribution, the generating function, \eqref{eq:degree distribution}, is
\begin{equation}\label{eq:G_UUU_ER}
G(x,z) = \sum_{j=0}^\infty \frac{\langle k\rangle ^j e^{-\langle k\rangle}}{j!} x^j z
\end{equation}

Inserting this expression in \eqref{eq:denominator}, the epidemic threshold can be expressed as (the full derivation is presented in \emph{Supplementary Information}, section 1)
\begin{equation}\label{eq:ER_UUU}\tag{ER-UUU}
T_c = \frac{1-T_{uv}}{\langle k\rangle + 1 - T_{uv}}
\end{equation}

Henceforth, to facilitate readability and unless otherwise stated, we provide expressions for the epidemic threshold in terms of the average transmission probability through intralinks, $T$, and the average transmission probability through interlinks, $T_{uv}$. Nevertheless, the thresholds can be easily rewritten in terms of $\beta_c$ in a straightforward way using \eqref{eq:T} and \eqref{eq:Tuv}. For this case, we can rewrite the above equation and explicitly express the value of $\beta_c$ as,
\begin{equation}\label{eq:ER_UUU_2}
\frac{\beta_c}{\mu} = \frac{1-T_{uv}}{\langle k \rangle}
\end{equation}

Note that if we set $\gamma=0$ in \eqref{eq:Tuv} so that the spreading from one layer to the other is completely removed, $T_{uv}=0$ and \eqref{eq:ER_UUU_2} is simplified to $\frac{\beta_c}{\mu} = \langle k \rangle^{-1}$, which is the classical value of the epidemic threshold in single layer ER networks \cite{Pastor-Satorras2001May}.

In a DUD network with nodes in both layers following a Poisson degree distribution, with the same average degree for both incoming and outgoing links, the generating function \eqref{eq:degree distribution} is
\begin{equation}\label{eq:G_DUD_ER}
G(x,y,z) = \sum_{j=0}^\infty \sum_{l=0}^\infty \frac{\langle k \rangle^j e^{-\langle k \rangle}}{j!} \frac{\langle k \rangle^l e^{-\langle k \rangle}}{l!} x^j y^l z
\end{equation}

Again, by inserting this expression in \eqref{eq:denominator} we obtain
\begin{equation}\label{eq:ER_DUD}\tag{ER-DUD}
T_c = \frac{1-T_{uv}}{\langle k\rangle}
\end{equation}

\begin{figure*}%
\begin{center}
\includegraphics[width=17.8cm]{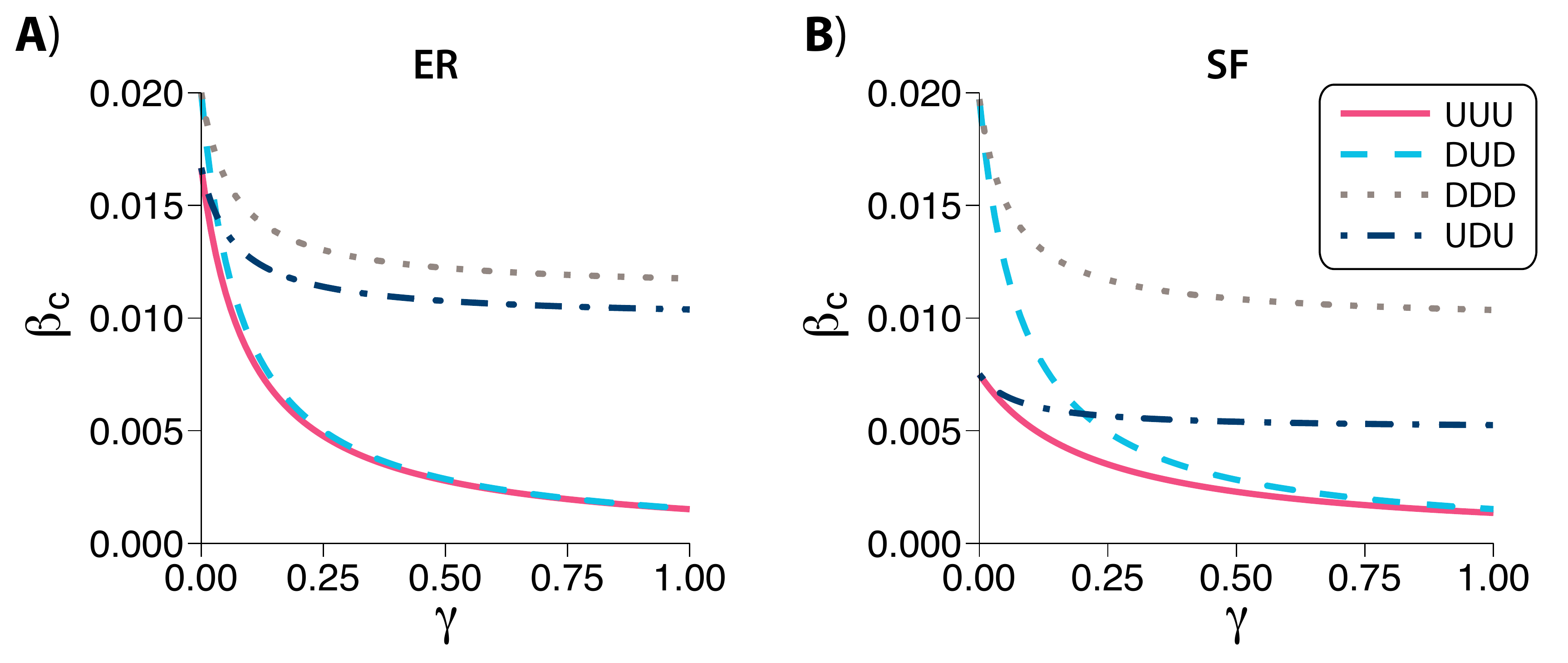}%
\end{center}
\caption{Comparison of the analytically derived epidemic thresholds for each network configuration UXU or DXD (X=U or D) and different degree distributions for the networks in the layers. A) ER networks with $\langle k \rangle = 6$ and $p=0.5$. B) SF networks with $k_\text{min}=3, \alpha=2.6$, resulting in the theoretical average degree $\langle k \rangle = 6.1$ and $p=0.5$.}
\label{fig:ER_SF}%
\end{figure*}

On the other hand, using the $p$-model previously described, the epidemic threshold in DDD configurations as a function of $p$ is
\begin{equation}\label{eq:ER_DDD}\tag{ER-DDD}
T_c = \frac{2}{\langle k \rangle ( 2 + m + \sqrt{m(m+8)})}
\end{equation}

\noindent
with $m = p(1-p)T_{uv}^2$ and in the UDU configuration is
\begin{equation}\label{eq:ER_UDU}\tag{ER-UDU}
T_c = \frac{2(1+\langle k \rangle) + m' - \sqrt{m' (4+8\langle k \rangle + m')}}{2((1+\langle k \rangle)^2-m'\langle k \rangle)}
\end{equation}

\noindent
with $m' = \langle k \rangle p(1-p)T_{uv}^2$. In figure \ref{fig:ER_SF}A we compare the behavior of these four configurations plotting $\beta_c$ as a function of $\gamma$.

\subsection{SF networks}

In SF networks the degree distribution follows a power-law of the form $P(k) \sim k^{-\alpha}$. Thus, the epidemic thresholds are
\begin{equation}\label{eq:SF_UUU}\tag{SF-UUU}
T_c = \frac{\langle k \rangle (1-T_{uv})}{\langle k^2\rangle (1-T_{uv}) + \langle k \rangle^2 T_{uv}}
\end{equation}

\noindent
for the UUU configuration,
\begin{equation}\label{eq:SF_DUD}\tag{SF-DUD}
T_c = \frac{1-T_{uv}}{\langle k \rangle}
\end{equation}

\noindent
for the DUD configuration,
\begin{equation}\label{eq:SF_DDD}\tag{SF-DDD}
T_c = \frac{2}{\langle k \rangle (2+m +\sqrt{m(m+8)}}
\end{equation}

\noindent
with $m=p(1-p)T_{uv}^2$ for the DDD configuration and
\begin{equation}\label{eq:SF_UDU}\tag{SF-UDU}
T_c = \frac{2\langle k^2\rangle\langle k \rangle + \langle k \rangle^2 \left(\langle k \rangle m - \sqrt{m(4\langle k^2\rangle + \langle k \rangle^2(4+m))}\right)}{2((\langle k^2 \rangle^2 - \langle k \rangle^4m)}
\end{equation}

\noindent
with $m=p(1-p)T_{uv}^2$  for the UDU configuration. The full derivation can be found in the \emph{Supplementary Information}, section 2. As in the ER case, the explicit dependence of $\beta_c$ with $\gamma$ is shown in figure \ref{fig:ER_SF}B.

\section{Discussion}

Our results show that directionality is a key factor in the spreading of epidemics in multiplex networks. Even more, these findings suggest that its effects cannot be trivially generalized as the consequences of changing the directionality of some links are completely different for Scale-Free and Erd\H{o}s-Rényi networks. In particular, in figure \ref{fig:UUU_DUD}A, we can see that for networks with $\langle k \rangle = 6$ the epidemic threshold is very similar in both UUU and DUD configurations. This effect is again seen for denser networks, $\langle k \rangle=12$, implying that it is the directionality of the interlinks, and not the one of the links contained within layers, the main driver of the epidemic in these networks. On the other hand, in figure \ref{fig:UUU_DUD}B we can see that this behavior is not replicated for SF networks. Certainly, there is a large difference between the curves of the UUU and DUD configurations, implying that the directionality of intralinks is much more important in this type of networks. In agreement with these observation, when the interlinks are those that are directed, we found the same difference between ER and SF networks. As can be observed in figure \ref{fig:ER_DDD_UDU}A, the evolution of the epidemic threshold as a function of $\gamma$ is again quantitatively similar for both DDD and UDU configurations. Conversely, in figure \ref{fig:ER_DDD_UDU}B, a difference between these configurations arises again for SF networks. Besides, in all the cases considered so far, figures \ref{fig:UUU_DUD} and \ref{fig:ER_DDD_UDU}, the epidemic threshold is always lower for those configurations with undirected links within the layers, compared to those in which those links are directed, given the same interlink directionality.

To get further insights into the mechanisms driving the behavior observed previously, we rely on the analytically derived thresholds and explore the evolution of $\beta_c$ as a function of $\gamma$ for the whole range of possible values of the latter parameter. Results are shown in figure \ref{fig:ER_SF}. In this case, we can see that the value of the epidemic threshold of the DUD configuration in SF networks tends to the value of the UUU case for large values of the spreading probability across layers, mimicking the behavior of ER networks. Thus, when $\gamma\rightarrow 1$ we reach the  state in which both networks exhibit the same properties, namely: (i) the epidemic threshold in DUD and UUU configurations is the same; (ii) XDX (X=U or D) configurations are almost not affected by the value of $\gamma$, except for the weakly couple regime (i.e., small values of $\gamma$). Hence, in general, one can conclude that the directionality (or lack of) of the interlinks is the main driver of the epidemic spreading process. The exception is the limit of small spreading from layer to layer, as in this scenario, the directionality of interlinks makes SF networks much more resilient, see the dashed-dotted line in \ref{fig:ER_SF}B. Altogether, the general conclusion is that directionality reduces the impact of disease spreading in multilayer systems.

\begin{figure*}%
\begin{center}
\includegraphics[width=17.8cm]{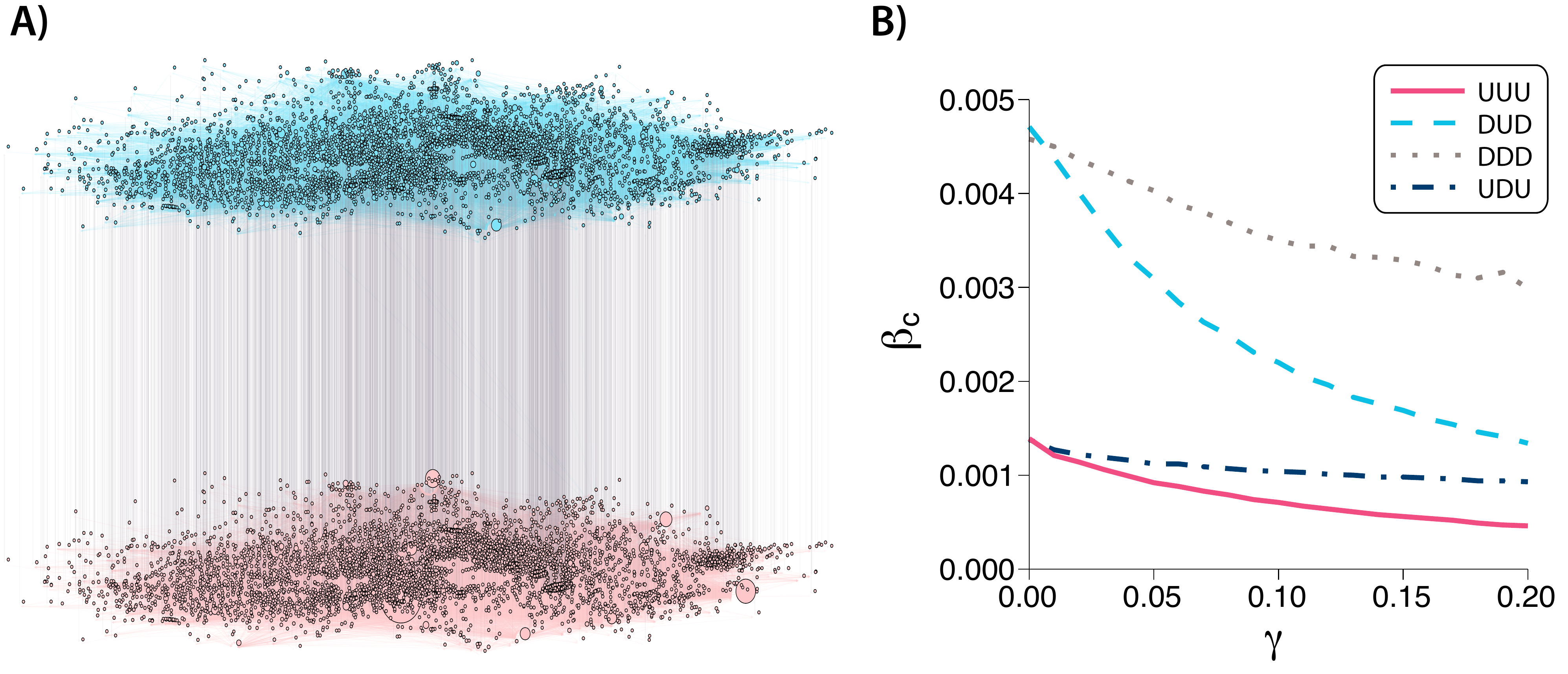}%
\end{center}
\caption{Epidemic threshold measured in a multiplex network composed by users of two different social platforms: friendfeed and twitter. The original network (A) has directed intralinks and undirected interlinks, thus it corresponds to the DUD configuration. Nevertheless, to explore the effects of directionality, the four configurations studied in this paper are considered (B). For those configurations with directed interlinks we used the $p$-model to generate them with, setting $p=0.5$. }
\label{fig:real}%
\end{figure*}

It is important to note that these results are not only relevant for the situations described in the introduction of this paper. First, because even though a system might be commonly presented as a monolayer network, it may be possible to detect different types of links in the network that would allow for the construction of a multiplex network. If this is done, as we have shown in this paper, the definition of the directionality of the interlinks is far from trivial as it can have dramatic consequences on the dynamics. In particular, the epidemic threshold can change by up to a factor of two depending on the directionality of the interlinks. Even more, these results are not restricted only to epidemic modeling, as these kind of diffusion processes can be applied to a broad range of systems. For example, the generating function approach has been proposed as a tool to identify influential spreaders in social networks \cite{Hu2018Jul}.

One particularly interesting and open challenge is to quantify the effects that the interplay between different social networks could have on spreading dynamics. The theoretical framework developed here is particularly suitable to study this and similar challenges related to the spreading of information in social networks. On the one hand, because social relations are, by default, directed: a user is not necessarily followed by her followings, i.e., social relations are not always reciprocal \cite{Mercken2010Jan}. On the other hand, disease-like models have been widely used to study information dissemination, or in other words, simple social contagion \cite{Weng2013Aug, Cozzo2013Nov}. We have analyzed the dependence of the epidemic threshold with the inter-spreading rate in a real social network composed by two layers, see figure \ref{fig:real}A. The first layer of the multilayer systems is made up by the directed set of interactions in a subset of users of the now defunct FriendFeed platform, whereas the second layer is defined by the directed set of interactions of those same users in Twitter. Even though this multiplex network originally corresponds to a DUD configuration, we have also explored the other possible configurations for the directionality of the links. Note that in contrast with the synthetic networks studied in the previous section, in this network the layers have different average degrees. In particular, the FriendFeed layer has 4,768 nodes and 29,501 directed links, resulting in an average out-degree of 6.19, and the Twitter layer is composed by 4,768 nodes and 40,168 directed links, with an average out-degree of 8.42. Nevertheless, their degree distributions are both heavy tailed, although the maximum degree in the FriendFeed network is much larger than in the Twitter network. For details on how this network was obtained, we refer the reader to the original source of the data \cite{Magnani2011}. 

The results, figure \ref{fig:real}B, confirm our findings for synthetic networks. In particular, for the range of $\gamma$ under consideration, the configurations with some directionality are always more resilient against the disease. These results would imply that information travels much more easily in undirected systems  than in directed systems. For instance, one could build up a directed multiplex network using Instagram and Twitter data, either in a DUD configuration if it is assumed that the likelihood of someone sharing the information from one platform to the other is independent of the source or in a DDD configuration if the likelihood of sending it from Instagram to Twitter is deemed to be different than from Twitter to Instagram. On the other hand, undirected social platforms such as Facebook and Whatsapp should be modeled using UDU or UUU configurations. According to our results, information would spread more easily through these platforms, which could be worrisome as they have recently been identified as one of the main sources of misinformation spreading \cite{nytimes}.
 
Lastly, it would be possible to build similar directed multiplex networks in transportation systems \cite{Aleta2017Mar}. In these systems, the interlinks can be modeled as undirected or directed, depending on the purpose of the study. If one is interested in taking into account the fact that, for example, a metro station can be overcrowded in the incoming direction but not in the outgoing direction, such as during the morning peak time, or the other way around, during the evening peak time, it would be necessary to use directed links. On the other hand, if congestion is not relevant for the study, those links could be regarded as undirected. 

In summary, we have developed a framework that allows studying disease-like processes in multilayer networks. This represents an important step towards the characterization of diffusion and spreading processes in interdependent multilevel complex systems. Our results show that directionality has a positive impact on the system's resistance to disease propagation and that the way in which interdependent (social) networks are coupled could determine their ability to spread information. Our results could be applied to a plethora of systems and show that more emphasis should be put in studying the role of interlinks in diffusion processes that take place on top of them.

\section{Materials and methods}
\subsection{Multilayer networks}
Multilayer networks are an extension of classical contact networks in which nodes are assigned to a given layer, $u$, and can be connected to nodes in the same layer or in other layers. As a result, it is possible to distinguish two types of links: intralayer links, which connect pairs of nodes in the same layer, and interlayer links, which connect pairs of nodes in different layers. This formulation is used to encode features that characterize the nodes or the links that would be otherwise hidden, such as different types of interactions in protein networks or the multiple transportation modes present in mass transit systems \cite{Aleta2018Oct}. In particular, in this work we focus on two layer directed multiplex networks. That is, networks composed by two layers in which links, either within layers or to other layers, can be directed. Furthermore, the term multiplex, in contrast to multilayer, implies that a node can only be connected to its counterpart in the other layer. In other words, it is not possible to have more than one link in each node going to the other layer \cite{Kivela2014Sep}.

\subsection{Stochastic simulations\label{mat:stochastic}}
The SIS dynamics is implemented on two layer multiplex networks with ER and SF topologies in the layers. In the simulations, all the nodes in the system are initially susceptible. The spreading starts when one node is set to the infectious state. Then, at each time step, each infected node spreads the disease through each of its links with probability $\beta$ if the link is contained in a layer and with probability $\gamma$ if the link connects nodes in different layers. Besides, each infected node recovers with probability $\mu$ at each time step. The simulation runs until a stationary state for the number of infected individuals is reached.

To determine the epidemic threshold we fix the value of $\gamma$ and run the simulation over multiple values of $\beta$, repeating $10^3$ times the simulation for each of those values. The minimum value of $\beta$ at which, on average, the number of infected individuals in the steady state is greater than one determines the value of the epidemic threshold, $\beta_c/\mu$. This procedure is then repeated for several values of $\gamma$ to obtain the dependency of $\beta_c$ with the spreading across layers. Lastly, this dependency is evaluated for $10^2$ realizations of each network considered in the study and their $\beta_c(\gamma)$ curves are averaged.

\begin{acknowledgements}X.W. acknowledges the support of the research program complexity in logistics (No. 439.16.107) from NOW. A.A. acknowledges the support of the FPI doctoral fellowship from MINECO and its mobility scheme. D.L. acknowledges support from the China Scholarship Council though a Ph.D fellowship. Y.M. acknowledges partial support from the Government of Arag\'on, Spain through a grant to the group FENOL (E36-17R), by MINECO and FEDER funds (grant FIS2017-87519-P) and by Intesa Sanpaolo Innovation Center. The funders had no role in study design, data collection and analysis, or preparation of the manuscript.\end{acknowledgements}

\bibliography{references}

\FloatBarrier
\newpage

\begin{center}
\textbf{\large Supplemental Materials: Directionality reduces the impact of epidemics in multilayer networks}
\end{center}
\setcounter{equation}{0}
\setcounter{figure}{0}
\setcounter{table}{0}
\setcounter{page}{1}
\makeatletter
\renewcommand{\theequation}{S\arabic{equation}}
\renewcommand{\thefigure}{S\arabic{figure}}
\renewcommand{\bibnumfmt}[1]{[S#1]}
\renewcommand{\citenumfont}[1]{S#1}

\section{Methods}
\subsection{General derivation of the epidemic threshold}

To understand the dynamical spreading of epidemics on directed multilayer networks, we mainly investigate how the epidemic threshold is influenced by the directionality between interacting individuals. In this section, we analytically derive, following the mathematical technique of generating functions, the epidemic threshold for SIS epidemic model on directed multilayer networks.

Consider a directed multiplex network consisting of two layers interconnected by interlinks. The directed contact between an infective individual to a susceptible individual can be within the same layer, or across different layers or a mixed of both. Depending on the directionality of links within layers and the directionality of links interconnecting different layers, we analyze all possible combinations which are (i) directed layers  and undirected interlinks, denoted as DUD, (ii) directed layers  and directed interlinks, denoted as DDD, (iii) undirected layers  and directed interlinks, denoted as UDU and (iv) undirected layers  and undirected interlinks, denoted as UUU.

For a general directed multilayer network, a node has an in-degree $ j $, out-degree $ k $ and inter-degree $ m $ with probability $ p_{jkm} $. The generating function for the degree distribution of a node is defined as
\begin{equation}\label{key}
G(x,y,z) = \sum_{j=0}^{\infty}\sum_{k=0}^{\infty}\sum_{m=0}^{\infty}p_{jkm}x^jy^kz^m
\end{equation}
\noindent
where $ G(1,1,1) = \sum_{j,k,m}p_{jkm}=1 $ satisfying the probability property.

Another quantity related to the nodal degree distribution is called the excess degree distribution, which is the distribution of degrees of nodes reached by following a randomly chosen link. The probability to reach a node is biased by nodal degrees because nodes with a higher degree have a higher probability to be chosen. The probability to reach a node by following the direction of a randomly chosen link, i.e., in-link of the reached node, is $ \frac{jp_{jkm}}{\sum_{j, k, m}p_{jkm}} $. The corresponding generating function for the excess in-degree $ j-1 $, out-degree $ k $ and inter-degree $ m $ reads
\begin{equation}\label{key}
H_d(x,y,z) =  \sum_{j=0}^{\infty}\sum_{k=0}^{\infty}\sum_{m=0}^{\infty}  \frac{jp_{jkm}}{\sum_{j=0}^{\infty}\sum_{k=0}^{\infty}\sum_{m=0}^{\infty} jp_{jkm}} x^{j-1}y^kz^m = \frac{G^{(1,0,0)}(x,y,z)}{G^{(1,0,0)}(1,1,1)}
\end{equation}
Analogously, the generating function for a node reached by following the reverse direction of a randomly chosen directed link, i.e., out-link of the reached node, follows
\begin{equation}\label{key}
H_{r}(x,y,z) =  \sum_{j=0}^{\infty}\sum_{k=0}^{\infty}\sum_{m=0}^{\infty}  \frac{kp_{jkm}}{\sum_{j=0}^{\infty}\sum_{k=0}^{\infty}\sum_{m=0}^{\infty} kp_{jkm}} x^{j}y^{k-1} z^m= \frac{G^{(0,1,0)}(x,y,z)}{G^{(0,1,0)}(1,1,1)}
\end{equation}
and similarly, the generating function for a node reached by following an undirected inter-link reads
\begin{equation}\label{key}
H_u(x,y,z) =  \sum_{j=0}^{\infty}\sum_{k=0}^{\infty}\sum_{m=0}^{\infty}  \frac{mp_{jkm}}{\sum_{j=0}^{\infty}\sum_{k=0}^{\infty}\sum_{m=0}^{\infty} mp_{jkm}} x^{j}y^{k} z^{m-1}= \frac{G^{(0,0,1)}(x,y,z)}{G^{(0,0,1)}(1,1,1)}
\end{equation}

To account for the probability of a link being infected by a disease that is transmitted from an infective individual to a susceptible individual, we further modify the generating functions. Denote $  T_i $, $ i \in {1, 2} $ as the average probability that a susceptible individual will be infected by an infectious individual in the same layer. Denote  $  T_{uv}$ as the average probability that an infectious individual from layer $ u$ will transmit the disease to a susceptible individual in layer $ v $. We omit the subscript of $ T_i $ when there is no ambiguous. The generating function for the distribution of the number of infected links of a randomly chosen node is obtained by incorporating the probability of disease transmission in the generating function of degree distribution, which reads
\begin{equation}\label{key}
\begin{split}
G(x,y,z,T,T_{uv} ) &=  \sum_{a, b, c}
\left[\sum_{j=a}^{\infty}\sum_{k=b}^{\infty}\sum_{m=c}^{\infty}p_{jkm}\binom{j}{a}T^a\left(1-T\right)^{j-a}\binom{k}{b}T^b\left(1-T\right)^{k-b}\binom{m}{c}T_{uv}^c(1-T_{uv})^{m-c}\right]x^ay^bz^c\\
& = \sum_{j=0}^{\infty}\sum_{k=0}^{\infty}\sum_{m=0}^{\infty}p_{jkm}\left(1-T+T x\right)^j\left(1-T+T y\right)^k\left(1-T_{uv}+T_{uv} z\right)^m\\
& = G(1-T+T x,1-T+T y,1-T_{uv}+T_{uv} z) 
\end{split}
\end{equation}

Analogously, we derive the generating functions, for the distribution of the number of infected links of a node reached by following a randomly chosen directed link in the designed direction, as
\begin{equation}\label{key}
H_d(x,y,z,T,T_{uv}) =  H_d(1-T+T x,1-T+T y,1-T_{uv}+T_{uv} z) 
\end{equation}
and similarly of a node reached by following a randomly chosen undirected inter-link, as 
\begin{equation}\label{key}
H_{u}(x,y,z,T,T_{uv}) =  H_u(1-T+T x,1-T+T y,1-T_{uv}+T_{uv}z) 
\end{equation}

A number of nodes can be infected starting from a single infected node within the directed multilayer network. Due to the randomness of disease spreading and the variability of contacts, the size of a disease outbreak is a random variable. To eventually determine the epidemic threshold, we first investigate the distribution of the size of an outbreak starting from a single infected node and its corresponding generating function. 

Denote $ \text{Pr}[S =s] $ as the probability of the size $ s $ of an outbreak starting from a single infected node. The generating function for the size distribution is defined as $ g(w,T,T_{uv})=  \sum_{s}\text{Pr}[S =s]w^s $. To solve the average size of an outbreak, we further define the generating function for the size of an outbreak starting from a node reached by a randomly chosen directed link in the designed direction, which denotes as $ h(w,T,T_{uv})=  \sum_{t}\text{Pr}[S =t]w^t  $. By adding subscript $ u $ or $ uv $ to the generating function $ h(w,T,T_{uv})$, we distinguish a randomly chosen link within a layer $ u$, $ u = 1, 2 $, and a randomly chosen interlink $ uv $ connecting layers $ u $ and $ v $.

\begin{figure}[!h]
	\centering
	\subfloat{
		\includegraphics[width=\textwidth]{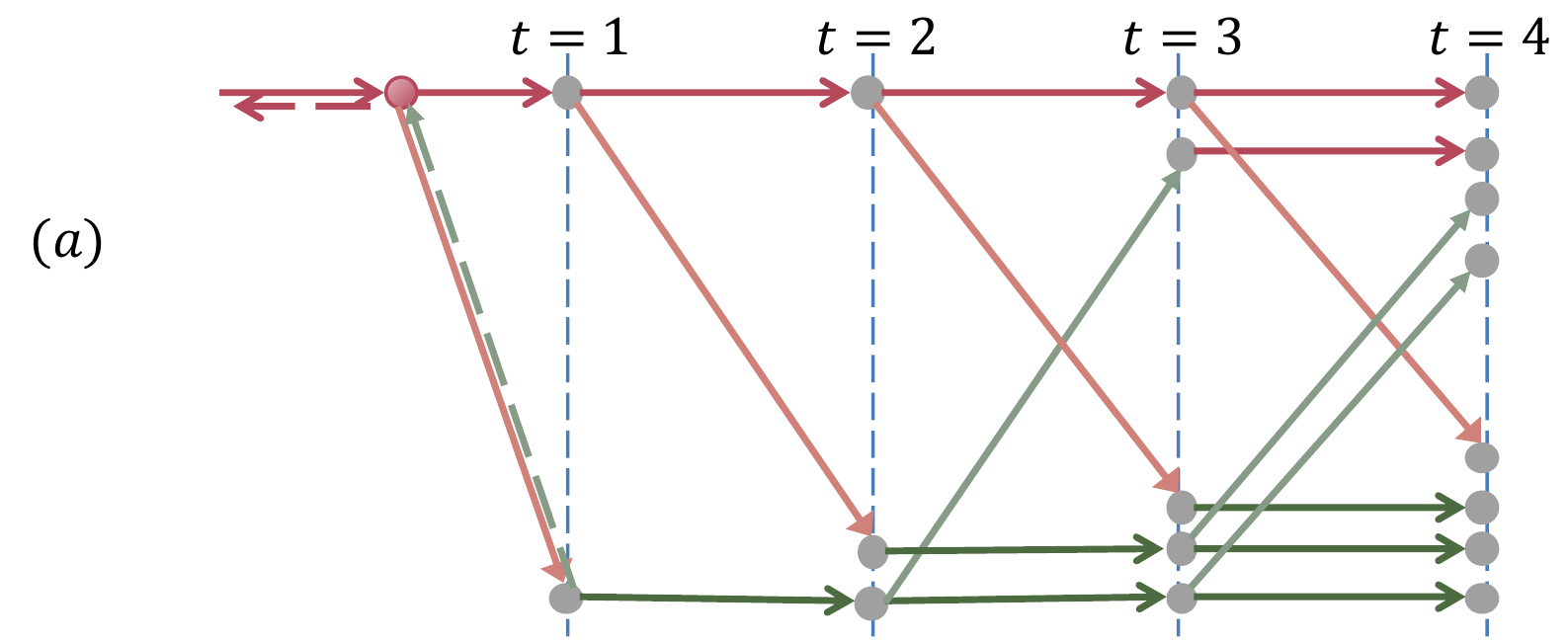}
		}\hfill
	\subfloat{
		\includegraphics[width=\textwidth]{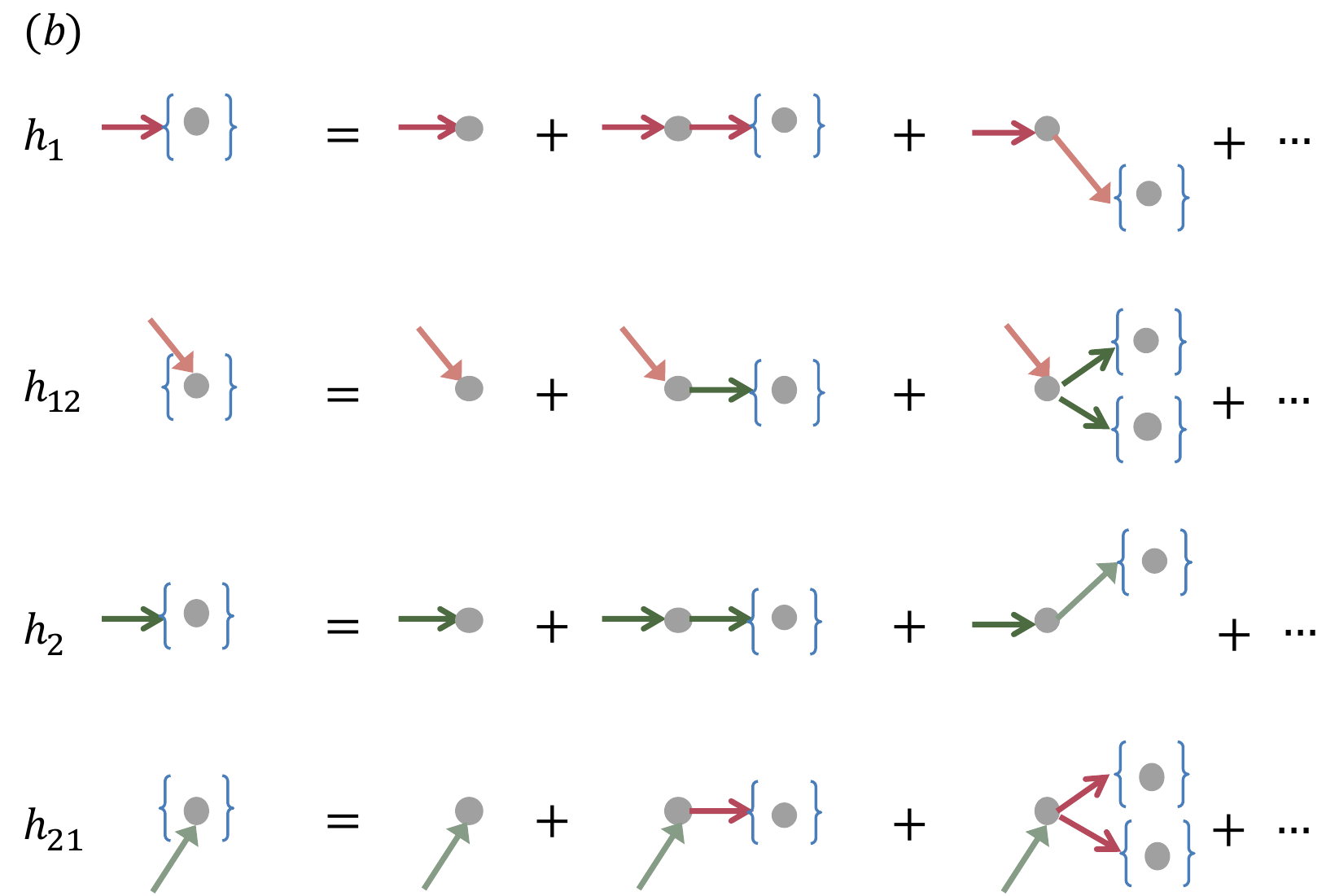}
		}
	\caption{Panel (a) shows the future transmission diagram starting from a single infected node reached by following the direction of a randomly chosen link. Solid lines represent the disease transmission on directed links and dashed lines depict the bidirectional disease transmission on undirected links. Panel (b) shows the recursive relation of generating functions for the size distribution of outbreaks by following four types of links which are (i) intralink in layer $ 1 $, (ii) interlink pointing from layer $ 1 $ to layer $ 2 $, (iii) intralink in layer 2 and (iv) interlink pointing from layer $ 2 $ to layer $ 1 $.}
	\label{fig_transimission_diagram}
\end{figure}

Starting from a single infected node reached by following a randomly chosen intra-link (links within layers), the possible ways of future transmission are: the disease spreads along an intra-link in the same layer, it spreads along an inter-link to the opposite layer, it spreads along two intra-links, it spreads along one intra-link and one inter-link, etc. The transmission diagram is shown in Fig. \ref{fig_transimission_diagram}(a). To account for all the transmission possibilities, we construct a recursive relation in the generating functions. Without loss of generality, we assume the disease spreading starting from an infected node in layer $ 1 $, the generating function satisfies a recursive relation
\begin{equation}\label{eq_generating_rnd_intralink}
h_1(w,T,T_{uv}) = w H_1(1,h_1(w,T,T_{uv}),h_{12}(w,T,T_{uv}),T,T_{uv})
\end{equation}
Generating function for the distribution of the size of an outbreak $ w $ along a randomly chosen interlink satisfies a recursive relation
\begin{equation}\label{key}
h_{12}(w,T,T_{uv})=wH_{12}(1,h_{2}(w,T,T_{uv}),h_{21}(w,T,T_{uv}),T,T_{uv})
\end{equation}
Analogously, the spreading in layer $ 2 $ itself satisfies a recursive relation
\begin{equation}\label{key}
h_{2}(w,T,T_{uv})=wH_{2}(1,h_{2}(w,T,T_{uv}),h_{21}(w,T,T_{uv}),T,T_{uv})
\end{equation}
and
\begin{equation}\label{key}
h_{21}(w,T,T_{uv})=wH_{21}(1,h_{1}(w,T,T_{uv}),h_{12}(w,T,T_{uv}),T,T_{uv})
\end{equation}
The recursive relation of the generating functions is shown in Fig. \ref{fig_transimission_diagram}(b). Similarly, generating function for the distribution of the size of an outbreak along a randomly chosen node in layer $ 1 $ follows
\begin{equation}\label{eq_generating_rnd_node}
g(w,T,T_{uv}) = wG(1,h_1(w,T,T_{uv}),h_{12}(w,T,T_{uv}),T,T_{uv})
\end{equation}
The average size $ E[S] $ of an outbreak starting from a randomly chosen node thus can be calculated by

\[
E[S]=\sum_{s = 1}^{N}s\text{Pr}[S =s] = \left.\frac{\text{d} g(w,T,T_{uv})}{\text{d} w}\right|_{w=1}
\]
Performing the derivative with respect to $ w $ on both sides of Eq. (\ref{eq_generating_rnd_intralink})-(\ref{eq_generating_rnd_node}), the derivatives for generating functions $ h_u $, $ h_{uv} $ and $ g $ read
\begin{equation}
\begin{gathered}
g^{'}(w,T,T_{uv})=G(1,h_{1},h_{12},T,T_{uv})+wG^{(0,1,0)}(1,h_{1},h_{12},T,T_{uv})h_{1}^{'}+wG^{(0,0,1)}(1,h_{1},h_{12},T,T_{uv})h_{12}^{'}
\\
h_{1}^{'}(w,T,T_{uv})=H_{1}(1,h_{1},h_{12},T,T_{uv})+wH_{1}^{(0,1,0)}(1,h_{1},h_{12},T,T_{uv})h_{1}^{'}+wH_{1}^{(0,0,1)}(1,h_{1},h_{12},T,T_{uv})h_{12}^{'}
\\
h_{12}^{'}(w,T,T_{uv})=H_{12}(1,h_{2},h_{21},T,T_{uv})+wH_{12}^{(0,1,0)}(1,h_{2},h_{21},T,T_{uv})h_{2}^{'}++wH_{12}^{(0,0,1)}(1,h_{2},h_{21},T,T_{uv})h_{21}^{'}
\\
h_{2}^{'}(w,T,T_{uv})=H_{2}(1,h_{2},h_{21},T,T_{uv})+wH_{2}^{(0,1,0)}(1,h_{2},h_{21},T,T_{uv})h_{2}^{'}+wH_{2}^{(0,0,1)}(1,h_{2},h_{21},T,T_{uv})h_{21}^{'}
\\
h_{21}^{'}(w,T,T_{uv})=H_{21}(1,h_{1},h_{12},T,T_{uv})+wH_{21}^{(0,1,0)}(1,h_{1},h_{12},T,T_{uv})h_{1}^{'}+wH_{21}^{(0,0,1)}(1,h_{1},h_{12},T,T_{uv})h_{12}^{'}
\end{gathered}
\end{equation}
For $ w=1 $, the derivatives of generating functions are simplified as
\begin{equation}
\begin{gathered}
g^{'}(1,T,T_{uv})=1+G^{(0,1,0)}h_{1}^{'}+G^{(0,0,1)}h_{12}^{'}
\\
h_{1}^{'}(1,T,T_{uv})=1+H_{1}^{(0,1,0)}h_{1}^{'}+H_{1}^{(0,0,1)}h_{12}^{'}
\\
h_{12}^{'}(1,T,T_{uv})=1+H_{12}^{(0,1,0)}h_{2}^{'}+H_{12}^{(0,0,1)}h_{21}^{'}
\\
h_{2}^{'}(1,T,T_{uv})=1+H_{2}^{(0,1,0)}h_{2}^{'}+H_{2}^{(0,0,1)}h_{21}^{'}
\\
h_{21}^{'}(1,T,T_{uv})=1+H_{21}^{(0,1,0)}h_{1}^{'}+H_{21}^{(0,0,1)}h_{12}^{'}
\end{gathered}
\end{equation}
where the arguments of a function in the right side of the equation are omitted for readability, for example $ G^{(0,1,0)}(1,h_{1},h_{12},T,T_{uv}) $  is denoted as $ G^{(0,1,0)} $.

Express the average size $E[s]$ of an outbreak in terms of the generating functions as
\[
\begin{split}
E[s]=g^{'}(1,T,T_{uv})=&1+\frac{{G^{(0,0,1)}\left(1+H_{12}^{(0,1,0)}h_{2}^{'}\right)}}{1-H_{12}^{(0,0,1)}H_{21}^{(0,0,1)}}\\
&+\frac{{\left(G^{(0,1,0)}\left(1-H_{12}^{(0,0,1)}H_{21}^{(0,0,1)}\right)+G^{(0,0,1)}H_{12}^{(0,0,1)}H_{21}^{(0,1,0)}\right)h_{1}^{'}}}{1-H_{12}^{(0,0,1)}H_{21}^{(0,0,1)}}
\end{split}
\]
where 
\begin{equation}\label{key}
h_{1}^{'}=\frac{{1+H_{1}^{(0,0,1)}+H_{1}^{(0,0,1)}H_{12}^{(0,0,1)}-H_{12}^{(0,0,1)}H_{21}^{(0,0,1)}+H_{1}^{(0,0,1)}H_{12}^{(0,1,0)}h_{2}^{'}}}{\left(1-H_{1}^{(0,1,0)}\right)\left(1-H_{12}^{(0,0,1)}H_{21}^{(0,0,1)}\right)-H_{1}^{(0,0,1)}H_{12}^{(0,0,1)}H_{21}^{(0,1,0)}}
\end{equation}
and
\begin{equation}\label{key}
h_{2}^{'}=\frac{{1+H_{2}^{(0,0,1)}+H_{2}^{(0,0,1)}H_{21}^{(0,0,1)}-H_{12}^{(0,0,1)}H_{21}^{(0,0,1)}+H_{2}^{(0,0,1)}H_{21}^{(0,1,0)}h_{1}^{'}}}{\left(1-H_{2}^{(0,1,0)}\right)\left(1-H_{12}^{(0,0,1)}H_{21}^{(0,0,1)}\right)-H_{2}^{(0,0,1)}H_{21}^{(0,0,1)}H_{12}^{(0,1,0)}}
\end{equation}
The expression for $ E[s] $ goes to infinity when the denominator equals zero, which characterizes a phase transition from small size of outbreaks with tree-like structure to the occurrence of large-scale outbreaks. Therefore, the critical equation that determines epidemic threshold reads
\begin{equation}\label{eq_cirtical_eq_epidemic_threshold}
\begin{split}
0=&\left[\left(1-H_{1}^{(0,1,0)}\right)\left(1-H_{12}^{(0,0,1)}H_{21}^{(0,0,1)}\right)-H_{1}^{(0,0,1)}H_{12}^{(0,0,1)}H_{21}^{(0,1,0)}\right]\\
&\left[\left(1-H_{2}^{(0,1,0)}\right)\left(1-H_{12}^{(0,0,1)}H_{21}^{(0,0,1)}\right)-H_{2}^{(0,0,1)}H_{21}^{(0,0,1)}H_{12}^{(0,1,0)}\right]-H_{1}^{(0,0,1)}H_{2}^{(0,0,1)}H_{12}^{(0,1,0)}H_{21}^{(0,1,0)}
\end{split}
\end{equation}

In the following subsections, we focus on deriving the specific epidemic threshold for the four configurations of DUD, UUU, DDD and UDU. The main approach is to determine generating functions in the critical equation \eqref{eq_cirtical_eq_epidemic_threshold}.
\subsection{Epidemic threshold for DUD}
Consider a directed multilayer network consisting of two directed graphs that are interconnected by undirected links. We employ Poisson degree distributions as an example to illustrate the derivation of the epidemic threshold. If both the in-degree and out-degree follow a Poisson distribution with the same average degree $ \langle k \rangle $, the generating function for the excess degree $ H_d $ follows
\begin{equation}\label{key}
H_{d}\left(x,y,z\right) = \frac{\sum_{j=0}^{\infty}\sum_{k=0}^{\infty}\frac{\langle k \rangle^je^{-\langle k \rangle}}{j!}\frac{\langle k \rangle^ke^{-\langle k \rangle}}{k!}jx^{j-1}y^{k}z}{ \langle k \rangle }
\end{equation}
from which we derive the partial derivative with respect to $ y $ evaluated at the point $ x=y=z=1 $ as
\begin{equation}\label{key}
H_{d}^{(0,1,0)}\left(1,1,1\right) = \langle k \rangle
\end{equation}
Since intralinks in the configuration of DUD are directed, the generation function $ H_1 $ for layer $ 1 $ is substituted by $ H_d $ which reads
\begin{equation}\label{key}
H_{1}^{(0,1,0)} = H_{d}^{(0,1,0)}\left(1,1,1,T,T_{uv}\right)=TH_{d}^{(0,1,0)}\left(1,1,1\right)
\end{equation}
The derivatives of the generating function $ H_1 $ for layer $ 1 $, and similarly $ H_2 $ for layer 2, thus follow
\begin{equation}\label{eq_gf_DUD_intra}
\begin{gathered}
H_{1}^{(0,1,0)}=H_{2}^{(0,1,0)}=T\langle k \rangle
\\
H_{12}^{(0,1,0)}=H_{21}^{(0,1,0)}=T\langle k \rangle
\end{gathered}
\end{equation}
As two layers of graphs are connected by undirected or bidirected interlinks, the disease thus can be transmitted with probability $ T_{uv} $ from layer $ 1 $ to layer $ 2 $ and, meanwhile, with probability $ T_{uv} $ to be transmitted from layer $ 2 $ to layer $ 1 $. The bidirectionality for disease transmission of undirected interlinks is reflected by the generating functions
\begin{equation}
\begin{gathered}
	H_{12}^{(0,1,0)}=H_u^{(0,1,0)}\left(1,1,1, T, T_{uv}\right) \\
	H_{12}^{(0,0,1)}= T_{uv} +H_u^{(0,0,1)}\left(1,1,1, T, T_{uv}\right)
\end{gathered}
\end{equation}
The extra added term $ T_{uv} $ incorporates the spreading from layer $ 2 $ to layer $ 1 $ due to the bi-directionality of an undirected interlink. With $ H_u^{(0,1,0)}\left(1,1,1, T, T_{uv}\right) =T\langle k \rangle $ for a Poisson degree distribution and $ H_u^{(0,0,1)}\left(1,1,1, T, T_{uv}\right)=0 $ for zero extra undirected interlink without the interlink we come along, we arrive at
\begin{equation}\label{eq_gf_DUD_inter}
\begin{gathered}
H_{12}^{(0,1,0)}=H_{21}^{(0,1,0)}=T\langle k \rangle
\\
H_{12}^{(0,0,1)}=H_{21}^{(0,0,1)}=T_{uv}
\end{gathered}
\end{equation}
Substituting generating functions (\ref{eq_gf_DUD_intra}) and (\ref{eq_gf_DUD_inter}) into the equation (\ref{eq_cirtical_eq_epidemic_threshold}) characterizing the critical point of phase transition, we derive the epidemic threshold for DUD as
\[
T_c=\frac{{1-T_{uv}}}{\langle k \rangle}
\]
\subsection{Epidemic threshold for UUU}
As a comparison with directed multilayer networks of directed layers and undirected interconnections, we investigate the undirected multilayer networks of undirected layers and undirected interlinks (UUU). In the case of UUU, both individuals within the same layer and across different layers can spread the disease to its neighbors and, in turn, can be infected by  its neighbors. Thus, generating functions for intra-layer spreading read
$ H_{1}^{(0,1,0)} =H_{2}^{({0,1,0})}=T+T\langle k \rangle  $ and for inter-layer spreading read $H_{1}^{(0,0,1)}=T_{uv}$ and $ H_{12}^{(0,0,1)}= T_{uv} $. All the generating functions needed to obtain epidemic threshold are
\begin{equation}
\begin{gathered}
H_{1}^{(0,1,0)}=H_{2}^{(0,1,0)}=T+T\langle k \rangle
\\
H_{12}^{(0,1,0)}=H_{21}^{(0,1,0)}=T\langle k \rangle
\\
H_{1}^{(0,0,1)}=H_{2}^{(0,0,1)}=T_{uv}
\\
H_{12}^{(0,0,1)}=  H_{21}^{(0,0,1)}=T_{uv}
\end{gathered}
\end{equation}
from which we obtain the epidemic threshold for UUU as
\begin{equation}\label{key}
T_c=\frac{{1-T_{uv}}}{\langle k \rangle+1-T_{uv}}
\end{equation}
\subsection{Epidemic threshold for DDD}
Moving from multilayer networks with undirected interconnections, we generalize the framework of generating functions to multilayer networks consisting of directed layers and directed interlinks. 

\subsubsection{$ p $ model to generate directed interlinks}
To generate directed interlinks, we introduce a $ p $ model, where an interlink is directed from layer $ 1 $ to layer $ 2 $ with probability $ p $ and with probability $ 1-p $ directed from layer $ 2 $ to layer $ 1 $.  The generation of a complete set of directed interlinks connecting different layers is determined by a single parameter $ p $. 

\begin{figure}[!h]
	\centering
	\includegraphics[width=0.65\textwidth]{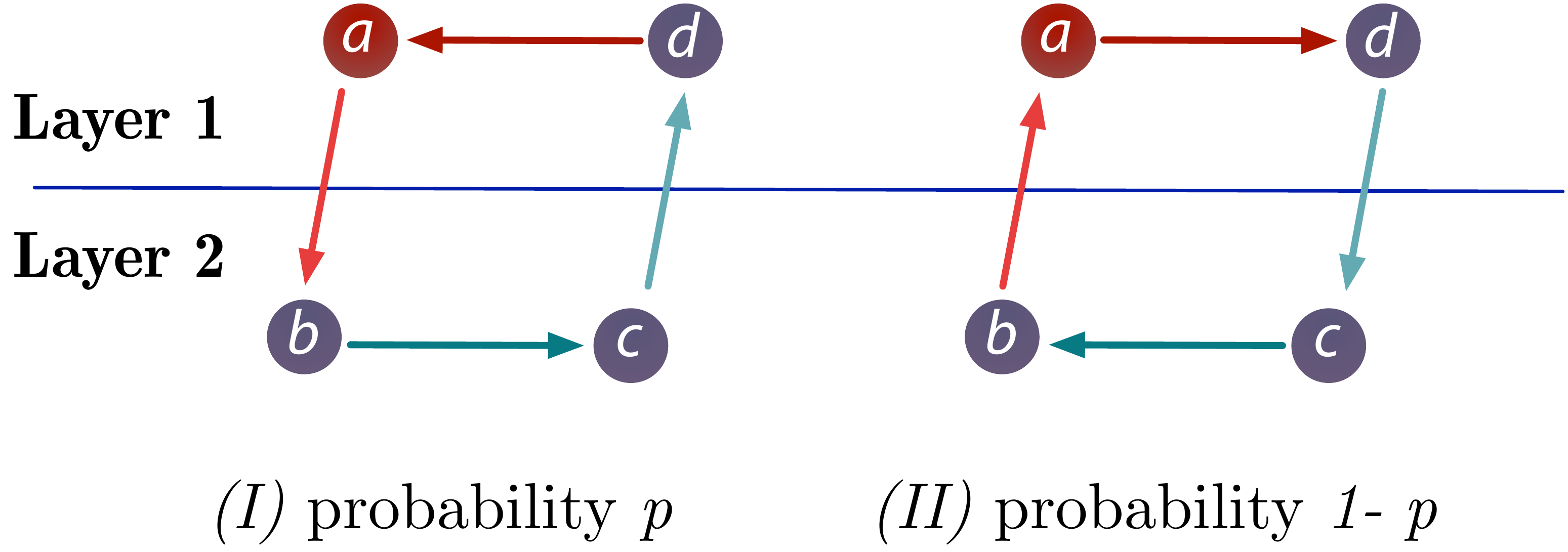}
	\caption{When spreading below the epidemic threshold, the disease, started from a randomly chosen infectious node, spreads in a tree-like structure. However, at the critical epidemic point, closed-loop forms and the expected size of outbreaks becomes infinity, which corresponds to the point of zero denominator in the corresponding generating function. In the case of DDD and UDU, two types of closed loops coexist, type I with probability $ p $ and type II with probability $ 1-p $. Starting from a randomly chosen infected node $ a $ in layer 1, the disease in type I spreads to the opposite layer 2 via the directed interlink $ a \rightarrow b $ from layer 1 to layer 2 (with probability $ p $) and loops back to the started node $ a $ via the directed path $ b \rightarrow c \rightarrow d \rightarrow a $. The spreading diagram in type II is the reverse of the diagram in type I, but with probability $ 1-p $ for a randomly chosen node to have an interlink directed from layer 2 to layer 1.} 
	\label{fig_closed_loop_DDD_UDU}
\end{figure}

An interlink generated by the $ p $ model has a single directionality between individuals from different layers. When a disease spreads from layer $ 1 $ to layer $ 2 $ with probability $ pT_{uv}   $, the backwards spreading from layer $ 2 $ to layer $ 1 $ via the same interlink is prohibited. The single directionality of disease transmission leads to $ H_{1}^{(0,0,1)}=pT_{uv}  $ and $ H_{12}^{(0,0,1)}=0 $.  When a disease spreading from layer $ 2 $ back to layer $ 1 $, it either forms a closed loop as type I or a closed loop as type II, where type I and type II are shown as Figure \ref{fig_closed_loop_DDD_UDU}. For closed loop of type I, we have $ H_{2}^{(0,0,1)}=(1-p)T_{uv} $. For the closed loop of type II, the spreading of disease loops back to the starting node once the interlink from layer 2 to layer 1 is occupied. The spreading via an intra-mediate link, characterized by $ H_{21}^{(0,1,0)} $ is therefore redundant.  Accordingly, we exclude $ H_{21}^{(0,1,0)} $ from $ H_{2}^{(0,0,1)} $ and arrive at $ H_{2}^{(0,0,1)}=\frac{{(1-p)T_{uv}}}{T\langle k \rangle} $.  Combined both type I and type II, we have $  H_{2}^{(0,0,1)} = (1-p)T_{uv}\left(1+\frac{{1}}{T\langle k \rangle}\right) $.

Together with the generating functions for disease spreading within layers, we have that
\begin{equation}\label{eq_generatings_DDD}
\begin{gathered}
H_{1}^{(0,1,0)}=H_{2}^{(0,1,0)}=T\langle k \rangle
\\
H_{12}^{(0,1,0)}=H_{21}^{(0,1,0)}=T\langle k \rangle
\\
H_{1}^{(0,0,1)}=pT_{uv} \quad \text{and} \quad H_{2}^{(0,0,1)}=(1-p)T_{uv}\left(1+\frac{{1}}{T\langle k \rangle}\right)
\\
H_{12}^{(0,0,1)}=0  \quad \text{and} \quad H_{21}^{(0,0,1)}=0
\end{gathered}
\end{equation}
from which the epidemic threshold for DDD reads
\begin{equation}\label{key}
T_c=\frac{{2}}{\langle k \rangle\left(2+m+\sqrt{m(m+8)}\right)}
\end{equation}
where $ m = p(1-p)T_{uv}^2 $
\subsubsection{$ pq $ model to generate directed interlinks}
As a single interlink generated by the $ p $ model has one and only one directionality, it might be limited to model scenarios of mixed directed and undirected contacts between layers. To allow for the coexistence of single directionality and bi-directionality of interlinks, we independently generate the directionality pointing from layer $ 1$ to layer $ 2$ with probability $ p $ and the directionality pointing from layer $ 2 $ to layer $ 1 $ with probability $ q $, which is termed as $ pq $ model. For the case of directed layers and directed interlinks generated by the $ pq $ model, we derive the epidemic threshold by substituting the following generating functions
\begin{equation}
\begin{gathered}
H_{1}^{(0,1,0)}=H_{2}^{(0,1,0)}=T\langle k \rangle
\\
H_{12}^{(0,1,0)}=H_{21}^{(0,1,0)}=T\langle k \rangle
\\
H_{1}^{(0,0,1)}=(p+q-pq)T_{uv} \quad \text{and} \quad H_{2}^{(0,0,1)}=(1-pq)T_{uv}
\\
H_{12}^{(0,0,1)}=(1-pq)T_{uv} \quad \text{and} \quad H_{21}^{(0,0,1)}=(p+q-pq)T_{uv}
\end{gathered}
\end{equation}
into (\ref{eq_cirtical_eq_epidemic_threshold}) which yields
\begin{equation}\label{key}
T_c=\frac{{1-T_{uv}\sqrt{{(p+q-pq)(1-pq)}}}}{\langle k \rangle}
\end{equation}
\subsection{Epidemic threshold for UDU}
In this section, we analyze the epidemic threshold for directed multilayer network of undirected layers and directed interlinks. 
\subsubsection{$ p $ model to generate directed interlinks}
In the case of UDU, the interlinks are generated by the same model as the network configuration of DDD. The generating functions for the spreading between layers remain the same. However, for disease spreading within layers, there are two types of future transmission which are spreading along excess links and spreading along the link that we come along. Correspondingly, the generating functions are modified as $ H_{1}^{(0,1,0)}=H_{2}^{(0,1,0)}=T+T\langle k \rangle $. All the generating functions to determine the epidemic threshold read  
\begin{equation}\label{eq_generatings_UDU}
\begin{gathered}
H_{1}^{(0,1,0)}=H_{2}^{(0,1,0)}=T+T\langle k \rangle
\\
H_{12}^{(0,1,0)}=H_{21}^{(0,1,0)}=T\langle k \rangle
\\
H_{1}^{(0,0,1)}=pT_{uv}  \quad \text{and} \quad   H_{2}^{(0,0,1)}=(1-p)T_{uv}\left(1+\frac{1}{T\langle k \rangle}\right)
\\
H_{12}^{(0,0,1)}=0  \ \text{and} \  H_{21}^{(0,0,1)}=0
\end{gathered}
\end{equation}
from which the epidemic threshold for UDU follows 
\begin{equation}
T_c = \frac{2\left(1+\langle k \rangle \right)+\langle k \rangle m-\sqrt{\langle k \rangle m\left(4+8\langle k \rangle+\langle k \rangle m\right)}}{2\left(\left(1+\langle k \rangle\right)^2-\langle k \rangle^2m\right)}
\end{equation}
where $ m = p(2-p)T_{uv} $
\subsubsection{$ pq $ model to generate directed interlinks}
The $ pq $ model generates a directed interlink from layer $ 1 $ to layer $ 2 $ with probability $ p+q-pq $. The reverse direction occurs with probability $ 1-pq $. The generating function for disease spreading from layer $ 1 $ to layer $ 2 $ reads $ H_{1}^{(0,0,1)}=(p+q-pq)T_{uv}   $ and the generating function for the disease to spread back follows $ H_{12}^{(0,0,1)}=(1-pq)T_{uv}  $. Together with the generating functions for intra-layer spreading, we have
\begin{equation}\label{eq_gfs_UDU_pq}
\begin{gathered}
H_{1}^{(0,1,0)}=H_{2}^{(0,1,0)}=T+T\langle k \rangle
\\
H_{12}^{(0,1,0)}=H_{21}^{(0,1,0)}=T\langle k \rangle
\\
H_{1}^{(0,0,1)}=(p+q-pq)T_{uv}  \quad \text{and} \quad   H_{2}^{(0,0,1)}=(1-pq)T_{uv}
\\
H_{12}^{(0,0,1)}=(1-pq)T_{uv}  \ \text{and} \  H_{21}^{(0,0,1)}=(p+q-pq)T_{uv}
\end{gathered}
\end{equation}
Substituting generating functions in (\ref{eq_gfs_UDU_pq}) into (\ref{eq_cirtical_eq_epidemic_threshold}) yields
\begin{equation}\label{key}
T_c=\frac{{1-T_{uv}\sqrt{{(p+q-pq)(1-pq)}}}}{1+\langle k \rangle-T_{uv}\sqrt{{(p+q-pq)(1-pq)}}}
\end{equation}
\subsection{Mapping rate to the transmission probability}
In the SIS epidemic model, a node at time $ t $ is either infected or healthy but susceptible to disease. An infected individual by a disease might recover from the disease or might spread to its direct neighbors. We assume both the recovering process and the spreading process are independent Poisson processes with rate $ \mu $ and $ \beta $, respectively. The time, denoted as $ \tau_i $, that an infected node $ i $ remains infected is a random variable, whose distribution follows an exponential distribution with rate $ \mu $.
 
The probability $ 1-T_{ij} $ that the disease will not transmit from an infected node $ i $ to a susceptible node $ j $ is $ e^{-\beta \tau_i} $. As $ \tau_i $ is a random variable, the probability $ T_{ij} $ of disease transmission is also a random variable. When assuming a homogeneous recovering rate for each node, the average of disease transmission probability between infected and susceptible individuals is the average over the distribution of infectious time, which follows 
\begin{equation}\label{key}
T = 1- \int_{0}^{\infty}e^{-\beta \tau} \mu e^{-\mu \tau}d\tau
\end{equation}
from which we obtain
\begin{equation}\label{key}
T = 1-\frac{\mu}{\beta+\mu}
\end{equation}
Analogously, the average transmission probability of individuals between different layers reads, given that the spreading rate between layers is $ \gamma $,
\begin{equation}\label{key}
T_{uv} = 1-\frac{\mu}{\gamma+\mu}
\end{equation}

\section{Additional results}
Although the derivation of the epidemic threshold has been so far done for Poisson degree distributions within each layer, this analytical framework can be generalized to different degree distributions. Specifically, we further present the generalization to directed multilayer networks consisting of scale-free networks with power-law degree distributions. 

For a power-law degree distribution with an exponential cutoff, the degree distribution is written as
\begin{equation}\label{key}
\text{Pr}[D = k] = Ck^{-\alpha}e^{-k/\kappa}
\end{equation}
where $ C=[\sum_{k_{\min}}^{k_{\max}}k^{-\alpha}e^{-k/\kappa}]^{-1} $ is a normalization constant to ensure $ \sum_{k_{\min}}^{k_{\max}} \text{Pr}[D=k]=1 $. The constant $ k_{\min} $ is the minimum degree and $ k_{\max} $ denotes the maximum degree and $ \kappa $ is a constant determining the cutoff.

Assume both the in- and out- degree in each layer are independently and identically power-law distributed. The generating function for the degree distribution of a node with in-degree $ i$, out-degree $j $ and inter-degree $ m $ reads
\begin{equation}\label{key}
G(x,y,z) = \sum_{j=0}^{\infty}\sum_{k=0}^{\infty}\sum_{m=0}^{\infty} \frac{j^{-\alpha}e^{-j/\kappa}k^{-\alpha}e^{-k/\kappa}p_m}{\sum_{j_{\min}}^{j_{\max}}j^{-\alpha}e^{-j/\kappa}\sum_{k_{\min}}^{k_{\max}}k^{-\alpha}e^{-k/\kappa}}x^jy^kz^m
\end{equation}
with $ p_m=1 $ to have an undirected interlink.
\subsection{UUU and DUD with scale-free layers}
The distributions of excess degree by following a randomly chosen link within a layer is accordingly modified by the power-law degree distribution. For the multiplex networks of UUU, the intra-links are undirected and thus the in-degree and out-degree within layers are indistinguishable. The generating function $ H_1 $ for the excess degree of a node reached by following a randomly chosen intra-link is modified as $ H_1(1,y,z) =  \sum_{k=0}^{\infty}\sum_{m=0}^{\infty} \frac{kp_{km}y^{k-1}z^m}{\langle k \rangle} $ from which
\begin{equation}\label{}
H_1^{(0,1,0)}(1,1,1)=\sum_{k=0}^{\infty}\sum_{m=0}^{\infty} \frac{k(k-1)p_{km}}{\langle k \rangle}=\frac{\langle k^2 \rangle -\langle k \rangle}{\langle k \rangle} 
\end{equation}
The ER graph with Poisson degree distribution is a special case where $ H_1^{(0,1,0)}(1,1,1)=\langle k \rangle $. Incorporating the transmission probability $ T $ of a disease on intra-links, we obtain $ H_1^{(0,1,0)}(1,1,1,T,T_{uv}) = T + T\frac{\langle k^2 \rangle -\langle k \rangle}{\langle k \rangle}  $, where the term $ T $ represents the backward spreading along the undirected intra-link that we came along and the term $ T\frac{\langle k^2 \rangle -\langle k \rangle}{\langle k \rangle} $ represents the spreading along the excess neighbors of the node that we reached. Incorporating the transmission probability $ T_{uv} $ of a disease on inter-links, we obtain $ H_1^{(0,0,1)}(1,1,1,T,T_{uv}) =T_{uv} $.

As the interconnection topology between different layers remains unchanged, the generating functions regarding interconnections are the same with generating functions of XUX with ER layers, i.e., $ H_{12}^{(0,0,1)}=H_{21}^{(0,0,1)}=T_{uv}$, $ H_{12}^{(0,1,0)}=H_{21}^{(0,1,0)}=T \langle k \rangle $. Substituting the following generating functions
\begin{equation}\label{eq_generatings_UUU_SF}
\begin{gathered}
H_{1}^{(0,1,0)}=H_{2}^{(0,1,0)}=T+T\frac{\langle k^2 \rangle -\langle k \rangle}{\langle k \rangle} 
\\
H_{12}^{(0,1,0)}=H_{21}^{(0,1,0)}=T\langle k \rangle
\\
H_{1}^{(0,0,1)}=H_{2}^{(0,0,1)}=T_{uv}
\\
H_{12}^{(0,0,1)}=  H_{21}^{(0,0,1)}=T_{uv}
\end{gathered}
\end{equation} 
into the critical equation (\ref{eq_cirtical_eq_epidemic_threshold}) yields the epidemic threshold for UUU with scale-free layers as 
\begin{equation}\label{eq_epidemic_threshold_SF_UUU}
T_c=\frac{\langle k \rangle\left(1-T_{uv}\right)}{\langle k^2 \rangle\left(1-T_{uv}\right)+\langle k \rangle^2T_{uv}} \tag{SF-UUU}
\end{equation} 

In the case of DUD with directed interlinks, the in-degree and out-degree are distinguishable and $ H_1^{(0,1,0)}(1,1,1)=H_2^{(0,1,0)}(1,1,1)= T\langle k \rangle $. The epidemic threshold for multiplex networks of DUD with scale-free layers remains unchanged with the ER layers, which reads
\begin{equation}\label{key}
T_c=\frac{{1-T_{uv}}}{\langle k \rangle} \tag{SF-DUD}
\end{equation}

\subsection{UDU and DDD with scale-free layers}
When the scale-free layers are interconnected by directed links, the generating functions $ H_{1}^{(0,0,1)} $ and $ H_{12}^{(0,0,1)} $ are modified to characterize the directionality of interlinks. As the interconnection patterns are unchanged from those of XDX with ER layers, the generating functions are referred to (\ref{eq_generatings_UDU}) for UDU and to (\ref{eq_generatings_DDD}) for DDD. Together with $ H_{1}^{(0,1,0)} $, $ H_{2}^{(0,1,0)} $ in (\ref{eq_generatings_UUU_SF}), we derive the epidemic threshold for UDU with scale-free layers as
\begin{equation}\label{key}\tag{SF-UDU}
T_{c}=\frac{2\langle k^2 \rangle \langle k \rangle+\langle k \rangle^2\left(\langle k \rangle m-\sqrt{m\left(4\langle k^2 \rangle+\langle k \rangle^2(4+m)\right)}\right)}{2\left(\langle k^2 \rangle^2-\langle k \rangle^4m\right)}
\end{equation}
where $ m = p(1-p)T_{uv} $.
With $ H_{1}^{(0,1,0)}=H_{2}^{(0,1,0)} = T\langle k \rangle $, the epidemic threshold for DDD with scale free-layers reads     
\begin{equation}\label{eq_epidemic_threshold_SF_DDD}
T_c=\frac{{2}}{\langle k \rangle\left(2+m+\sqrt{m(m+8)}\right)} \tag{SF-DDD}
\end{equation}
where $ m = p(1-p)T_{uv}^2 $.
\newpage

\begin{figure}[!h]
	\centering
	\includegraphics[width=\textwidth]{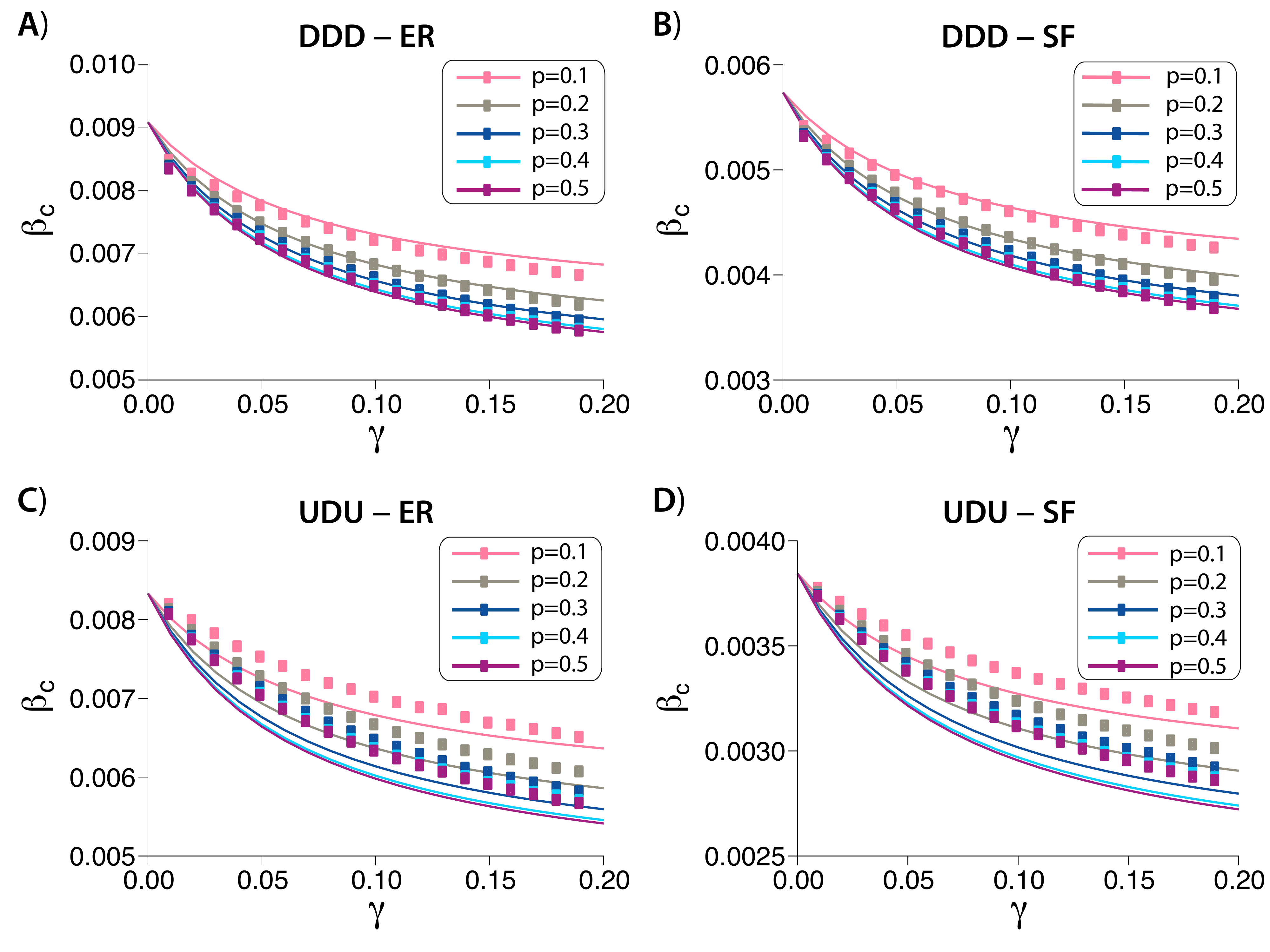}
	\caption{Epidemic threshold for two-layered multiplex networks for different values of $p$ with: (a) directed layers with ER degree distribution and and directed interlinks; (b) directed layers with SF degree distribution and directed interlinks; (c) undirected layers with ER distribution and directed interlinks; (d) undirected layers with SF distribution and directed interlinks. In both ER cases the average degree is 12 and in the SF cases the minimum degree is 10 and the exponent is 2.8, resulting in an average degree of 18.50.}
	\label{fig_critical_threshold_four_ER_network_configuration_pqinterlink_model}
\end{figure}

\end{document}